\documentstyle[12pt]{article}

\newcounter{eq}
\newcounter{sc}

\def\overleftrightarrow#1{\vbox{\ialign{##\crcr
 $\leftrightarrow$\crcr\noalign{\kern-1pt\nointerlineskip}
 $\hfil\displaystyle{#1}\hfil$\crcr}}}

\newlength{\minitwocolumn}
\begin{document}

\begin{flushright}
DPUR/TH/27\\
October, 2011\\
\end{flushright}
\vspace{20pt}

\pagestyle{empty}
\baselineskip15pt

\begin{center}
{\large\bf Superluminal Neutrinos from Gauge Field
\vskip 1mm }

\vspace{20mm}
Ichiro Oda \footnote{E-mail address:\ ioda@phys.u-ryukyu.ac.jp}
and Hajime Taira

\vspace{5mm}
           Department of Physics, Faculty of Science, University of the 
           Ryukyus,\\
           Nishihara, Okinawa 903-0213, Japan.\\

\end{center}


\vspace{5mm}
\begin{abstract}
We consider a possibility that the recent OPERA results on neutrinos' 
superluminality could be caused by a local effect of a new gauge field sourced
by the earth. If neutrinos couple to this gauge field via a gauge-invariant but
non-renormalizable interaction, the coupling effectively changes a background metric, 
thereby leading to superluminality of neutrinos. This possibility naturally might 
explain why neutrinos from CERN CNGS beam to Gran Sasso Laboratory become superluminal 
while those from SN1987A to Earth become subluminal.    
\end{abstract}

\newpage
\pagestyle{plain}
\pagenumbering{arabic}


The OPERA experiment has recently announced a sensational result,
that is, it was observed that neutrinos travel faster than the speed of 
light \cite{OPERA}. Although this remarkable discovery needs further experimental scrutiny 
in future, in the theoretical side it might call for consideration beyond 
standard physics, such as violation of the Lorentz invariance or extra-dimensional 
scenarios.

However, it is known that the breakdown of relativity could give us many 
theoretical pathologies. For instance, one could construct perpetuum mobile 
devices employing black holes which would have different horizons for different 
particle species. Moreover, one should recall that the Lorentz invariance
provides a strong constraint on the selection of actions in quantum field
theories. Indeed, under the situation of the absence of the Lorentz invariance,
one has no idea how to determine actions. It is the manifest Lorentz invariance 
that supports the success of Standard Model at the fundamental level.

Thus, it is worth persuing some possibilities for explaining the OPERA's results
as well as SN1987A ones \cite{SN} at the same time from a Lorentz-invariant theory. 
Actually, there have appeared such theories based on the Lorentz invariance 
where the cause of superluminal propagation of neutrinos is seeked in our earth
itself and the gravitational field. Put differently, the main idea is that superluminal
neutrinos are local effects triggered by a new massive spin-2 tensor field \cite{Dvali, Iorio}
and a new spin-0 scalar field of the Galileon type \cite{Kehagias} which exist
in the neighborhood of our earth. Then it is a natural question to ask ourselves 
whether or not the similar theories could be constructed in terms of a new spin-1 gauge field 
since the gauge field is on an equal footing with the tensor and scalar fields as elementary
particles. 
In this article, we will see that this is the case, and we shall make an explicit theory 
using the magnetic field sourced by the earth.

We shall begin with the Lagrangian density of our theory : \footnote{We make use of the metric
signature $\eta_{\mu\nu} = diag ( +, -, -, -)$.}
\begin{eqnarray}
{\cal{L}} = \bar{\psi} ( i \gamma^\mu \overleftrightarrow{\partial}_\mu - m ) \psi 
- i \frac{1}{M_*^4} F_\mu \ ^\alpha F_{\alpha\nu} \bar{\psi} \gamma^\mu 
\overleftrightarrow{\partial}^\nu \psi
- \frac{1}{4} F_{\mu\nu} F^{\mu\nu},
\label{Original Action}
\end{eqnarray}
where $M_*$ is a mass scale which controls the strength of the coupling and
the gauge field strength $F_{\mu\nu}$ is defined by $F_{\mu\nu} = \partial_\mu A_\nu
- \partial_\nu A_\mu$ as usual.  Moreover, we have defined 
$\bar{\psi} \gamma^\mu \overleftrightarrow{\partial}_\mu \psi 
\equiv \frac{1}{2} ( \bar{\psi} \gamma^\mu \partial_\mu \psi 
- \partial_\mu \bar{\psi} \gamma^\mu \psi )$. Because of this definition,
the Lagrangian density (\ref{Original Action}) is manifestly unitary. 
The fields $A_\mu$ and $\psi$ denote a new spin-1 gauge 
field and the spin-1/2 neutrino field, respectively. For simplicity of presentation,
we shall set mass of neutrinos to be vanishing, i.e., $m = 0$ in what follows.
Note that compared to the conventional system of spinor plus gauge fields, as a feature
of the Lagrangian density (\ref{Original Action}) there
exists a non-renormalizable but gauge-invariant coupling (therefore, unitary without
a ghost), which provides the following effective metric that neutrinos see
\begin{eqnarray}
g^{(\nu)}_{\mu\nu} = \eta_{\mu\nu} - \frac{1}{M_*^4} F_\mu \ ^\alpha F_{\alpha\nu}.
\label{E-metric}
\end{eqnarray}
Incidentally, one could pick up a linear coupling of the gauge field strength
to neutrinos, that is, $- i \frac{1}{M_*^2} F_{\mu\nu} \bar{\psi} \gamma^\mu \partial^\nu \psi$, 
but it is easy to show that this simpler coupling does not yield an effective
metric since $F_{\mu\nu}$ is an anti-symmetric tensor rather than a symmetric one. 
 
Next, let us consider what configuration the new gauge field takes in the local neighborhood
of our earth. To do that, it is interesting to proceed by analogy to the real
electro-magnetic field existing on the earth although this consideration is not always 
necessary. From the physical point of view, there is not so much electric charge 
on the earth since the excessive positive or negative electric charge 
would be almost neutralized on the earth (Of course, there is a feeble electric current 
on the earth, which is called "earth current" or "telluric current" \cite{GSI}). 
Indeed, by following the similar line of argument mentioned later, it is of interest that one can 
prove that the static electric potential $A_t = \frac{\alpha}{r}$ with $\alpha$ being a constant 
does not lead to superluminal propagation of neutrinos.

On the other hand, we are empirically familiar with the fact that there is the earth's magnetic 
field, which is dubbed "geomagnetism" and is generated in the fluid outer core by a self-exciting 
dynamo process. The earth's magnetic field is known to be approximately described by a small
magnetic dipole sitting at the center of the earth. Let us assume for simplicity that the magnetic field 
takes a value in the direction of the $z$ axis \footnote{Geomagnetism is not parallel to the surface 
of the earth except region near the equator.} and ask if neutrions propagate superluminally 
because of an effective metric which is induced by the magnetic field. In this case, 
the magnetic field is given by
\begin{eqnarray}
F_{12} \equiv B_z = \frac{\alpha}{r^2},
\label{M-field}
\end{eqnarray}
where $\alpha$ is a constant and its sign is not important since it appears in a quadratic form in
the formulae below, and $r$ is the distance from the center of the earth. 
Then, the equation of motion for the spinor field reads 
\begin{eqnarray}
\gamma^0 \partial_t \psi + \gamma^1 ( 1 + \frac{1}{M_*^4} \frac{\alpha^2}{r^4} ) \partial_x \psi 
+ \gamma^2 ( 1 + \frac{1}{M_*^4} \frac{\alpha^2}{r^4} ) \partial_y \psi
+ \gamma^3 \partial_z \psi -  \frac{2 \alpha^2}{M_*^4 r^6} ( x \gamma^1 + y \gamma^2 ) \psi = 0,
\label{Eq-motion}
\end{eqnarray}
where the last term denotes a field-dependent mass term which is neglected in what
follows since the size is very tiny. \footnote{Using (\ref{Beta}) and $R_E = 6.4 \times 10^6 m$, 
this field-dependent mass is approximately $\hbar c \frac{\alpha^2}{M_*^4 R_E^5}
= 10^{-5} \hbar c \frac{1}{R_E} = 10^{-16} eV$.}

Comparing this with the Dirac equation in a curved space-time 
\begin{eqnarray}
e^\mu_A \gamma^A \partial_\mu \psi = 0,
\label{Dirac}
\end{eqnarray}
where $e^\mu_A$ is the vierbein, and $\mu$ and $A$ are curved and flat
indices, respectively, we can easily read off non-vanishing components
of the vierbein
\begin{eqnarray}
e^t_0 = e^z_3 = 1, \quad e^x_1 = e^y_2 = 1 + \frac{1}{M_*^4} \frac{\alpha^2}{r^4}.
\label{Vierbein}
\end{eqnarray}

Next, using the relation between the metric tensor and the vierbein 
\begin{eqnarray}
g^{\mu\nu} = e^\mu_A e^\nu_B \eta^{AB},
\label{M-V relation}
\end{eqnarray}
the effective metric takes the form
\begin{eqnarray}
g^{tt} = 1, \quad g^{xx} = g^{yy} = - ( 1 + \frac{1}{M_*^4} \frac{\alpha^2}{r^4} )^2,
\quad g^{zz} = - 1.
\label{Metric}
\end{eqnarray}
Accordingly, the effective space-time on which neutrinos propagate
has the line element
\begin{eqnarray}
ds^2 = dt^2 - \frac{1}{(1 + \frac{1}{M_*^4} \frac{\alpha^2}{r^4})^2} 
( dx^2 + dy^2) - dz^2.
\label{Line}
\end{eqnarray}

Then, along the null trajectory (since neutrinos are assumed to be massless)
with the fixed $y$ and $z$ directions, the velocity of neutrinos is given by
\begin{eqnarray}
v \equiv  \frac{dx}{dt} = 1 + \frac{1}{M_*^4} \frac{\alpha^2}{r^4}.
\label{Velocity}
\end{eqnarray}
The result of OPERA gives us the condition for the dimensionless quantity $\beta$
\begin{eqnarray}
\beta \equiv  \frac{v - c}{c} = \frac{1}{M_*^4} \frac{\alpha^2}{r^4} = 10^{-5}.
\label{Beta}
\end{eqnarray}

Now we have to specify the coupling of the new gauge field $A_\mu$ to other 
particles in Standard Model in order to gain information on the mass scale
$M_*$. Let us assume that magnetic charge of the earth sources the gauge field
and creates a local magnetic field. This field modifies a local gravitational
background which neutrinos see and as a result the velocity of neutrinos
becomes larger than that of light. Thus, we assume that the coupling to the
rest of particles is given by $A_\mu J^\mu$ where $J^\mu$ is the electro-magnetic
current which does not include neutrinos.

With this assumption, the magnetic field on the earth is about $0.5 \times 10^{-4}
$ Tesla, so we have the relation
\begin{eqnarray}
\frac{\alpha}{R_E^2} = 0.5 \times 10^{-4},
\label{Magne}
\end{eqnarray}
where $R_E$ is the radius of the earth and takes the value $R_E = 6.4 \times 10^6 m$.
Recovering dimensional factors, it turns out that the mass scale is described as
\begin{eqnarray}
M_*^4 = \frac{\hbar^3}{\mu_0 c^5} (\frac{\alpha}{R_E^2})^2 \times 10^5,
\label{Mass-scale}
\end{eqnarray}
where $\mu_0$ denotes magnetic permeability of the vacuum. From this expression,
we arrive at the value of the mass scale in the theory at hand
\begin{eqnarray}
M_* \approx 10^{-6} MeV.
\label{M_*}
\end{eqnarray}
 
This scale is so low that it is tempting to identify the new gauge field
with the conventional electro-magnetic field, but to do so would need
more investigation in future. Here we shall assume that this conjecture 
holds and move on to the issue of subluminal neutrinos from SN1987A to Earth.
If the new gauge field were the familiar electro-magnetic field, the speed of
light would remain the velocity of light even if the velocity of neutrions
becomes superluminal by the mechanism mentioned thus far. In this sense, our
theory might give a nice resolution to the problem why neutrinos from 
CERN CNGS beam to Gran Sasso Laboratory become superluminal 
while those from SN1987A to Earth become subluminal.       

Finally, let us comment on the relation between the present theory and
Cohen-Glashow one \cite{Cohen}. The superluminal interpretation of the
OPERA results has been recently refuted theoretically by Cherenkov-like
radiation \cite{Cohen}. However, more recently three
different resolutions to the Cohen-Glashow result have been proposed
by three groups \cite{Aref'eva, Oda, Evslin}. In particular, it has been
shown in the last article \cite{Evslin} that if the velocity of an electron 
is equal to that of a neutrino, the threshold neutrino energy for the Cherenkov-like
radiation becomes infinite and the process will not occur kinematically
since the threshold neutrino energy is given by $E_\nu = \frac{2 m_e}
{\sqrt{v_\nu^2 - v_e^2}}$. In our work, when we generalize the spinor 
field $\psi$ to the SU(2) doublet $L = ( \nu_e, e )^T$ where $\nu_e$
and $e$ are respectively the neutrino spinor field and electron one,
the velocity of the neutrino and the electron takes the same
value owing to the mechanism of our theory (Recall that neutrino 
oscillation requires the three kinds of neutrinos to have the almost same 
velocity).
Thus, our theory amounts to realizing the resolution proposed by Evslin \cite{Evslin}.
Furthermore, one might fear that the interaction term $- i \frac{1}{M_*^4} F_\mu \ ^\alpha 
F_{\alpha\nu} \bar{\psi} \gamma^\mu \overleftrightarrow{\partial}^\nu \psi$ 
in (\ref{Original Action}) would induce the bremsstrahlung of photons, for instance.
Since this term is non-renormalizable and makes sense of only for energies less than 
the mass scale $M_*$, one can show that such the bremsstrahlung of photons does not
make a large contribution as follows: For simplicity, let us calculate the tree 
amplitudes of $\psi A^2 \rightarrow \psi A^2$ at the lowest order whose result takes 
the form $ m^3 (\frac{1}{M_*^4})^2 \frac{(p^3)^2}{p}|_{p < E} \sim \frac{m^3 E^5}{M_*^8}$.
Since we assume that $E < M_*$ and the mass of a neutrino is small, this contribution is 
rather small. Moreover, one finds that the higher-order tree amplitudes become smaller than
this lowest one, so we can safely ignore such the bremsstrahlung as long as we stay in 
the weak coupling region and we assume the mass of a neutrino to be small.

\begin{flushleft}
{\bf Acknowledgements}
\end{flushleft}

This work (I.O.) is supported in part by the Grant-in-Aid for Scientific 
Research (C) No. 22540287 from the Japan Ministry of Education, Culture, 
Sports, Science and Technology.



\begin{thebibliography}{99}


\bibitem{OPERA}
OPERA Collaboration, T. Adams et al., {"Measurement of the neutrino velocity
with the OPERA detector in the CNGS beam", 
arXiv:1109.4897 [hep-ex].}

\bibitem{SN}
K. Hirata et al., {"Observation of a neutrino burst from the supernova
SN1987A", Phys. Rev. Lett. 58 (1987) 1490.}

\bibitem{Dvali}
G. Dvali and A. Vikman, {"Price for environment neutrino-superluminality", 
arXiv:1109.5685 [hep-ph].} 

\bibitem{Iorio}
L. Iorio, {"Environmental fifth-force hypothesis for the OPERA superluminal
neutrino phenomenology: constraints from orbital motions around the Earth", 
arXiv:1109.6249 [gr-qc].} 

\bibitem{Kehagias}
A. Kehagias, {"Relativistic superluminal neutrinos", 
arXiv:1109.6312 [hep-ph].} 

\bibitem{GSI}
The Geospatial Information Authority of Japan (GSI), {ULR: http://vldb.gsi.go.jp/
sokuchi/geomag/index.html (in Japanese).} 

\bibitem{Cohen}
A. G. Cohen and S. L. Glashow, {"New Constraints on Neutrino Velocities",
arXiv:1109.6562 [hep-ph].} 

\bibitem{Aref'eva}
I.Ya. Aref'eva and I. V. Volovich, {"Superluminal Dark Neutrinos",
arXiv:1110.0456 [hep-ph]} 

\bibitem{Oda}
I. Oda and H. Taira, {"A Resolution to Cherenkov-like Radiation of OPERA Neutrinos",
arXiv:1110.6571 [hep-ph].} 

\bibitem{Evslin}
J. Evslin, {"Challenges Confronting Superluminal Neutrinos Models",
arXiv:1111.0733 [hep-ph].} 


\end{thebibliography}
\end{document}